# Chemical segregation in $Ge_2Sb_2Te_5$ thin films during in-situ heating

Shalini Tripathi[1], Paul Kotula[2], Manish Singh[1], Chanchal Ghosh[1], Gokhan Bakan[3], Helena Silva[1], Barry Carter[4,5]

[1]Electrical & Computer Engineering, University of Connecticut, Storrs, CT 06269
[2]Materials Science & Engineering Center, Sandia National Lab, Albuquerque, NM 87123
[3]The National Graphene Institute, University of Manchester, Manchester, M13 9PL, UK
[4]Chemical & Biomolecular Engineering, University of Connecticut, Storrs, CT 06269
[5]Center for Integrated Nanotechnologies (CINT), Sandia National Lab, Albuquerque, NM 87123

## Abstract

Germanium antimony telluride has been the most used and studied phase-change material for electronic memory due to its suitable crystallization temperature, amorphous to crystalline resistance contrast, and stability of the amorphous phase. In this work, the segregation of Ge in a $Ge_2Sb_2Te_5$ film of 30 nm thickness during heating inside the transmission electron microscope was observed and characterized. The $Ge_2Sb_2Te_5$ film was deposited using sputtering on a Protochips Fusion holder and left uncapped in atmosphere for about four months. Oxygen incorporated within the film played a significant role in the chemical segregation observed which resulted in amorphous Ge-O grain boundaries and Sb and Te rich crystalline domains. Such composition changes can occur when the phase-change material interfaces insulating oxide layers in an integrated device and would significantly impact its electrical and thermal properties.

Phase-change memory (PCM) is a new technology for non-volatile electronic memory, significantly faster than flash memory[1-3]. Extensive work on the crystallization and amorphization properties of phase-change materials has been carried out to clarify the functional physical properties of suitable materials and explore their phase transformation dynamics. In addition to the phase transformation, chemical segregation and oxidation of the material are important issues that must be understood for proper integration of PCM devices. The most commonly used material for PCM has been $Ge_2Sb_2Te_5$, which can exist as metastable amorphous and crystalline *fcc* phases, and as stable crystalline *hexagonal* phase[4,5,6]. Over the last decades $Ge_2Sb_2Te_5$ has attracted the attention of materials researchers as the potential candidate for PCM for the future generation of non-volatile memory applications[7-9]. However, most of the reports documented in these materials concern measurements of their resistivity and their functional behavior; studies on microscopic characterization of their microstructural and microchemical nature are relatively sparse. In the literature, these types of materials are reported to be prone to oxidation, especially Ge, resulting in the formation of germanium oxide.

It has been reported that phase segregation takes place at the interface between the GST and electrodes, limiting device endurance[10]. It has also been observed in endurance tests of $Ge_2Sb_2Te_5$ devices that Ge segregates towards material interfaces and can oxidize there[11]. Studies on oxygen-incorporated $Ge_2Sb_2Te_5$ films showed an increase in the amorphous-fcc phase transition and formation of non-stoichiometric GeO and phase separation into $Sb_2O_3$ and $Sb_2Te_3$[12].

Oxidation of the $Ge_2Sb_2Te_5$ films always has an effect on their phase transformation behavior and the reaction kinetics with thermal treatment[13-15]. This is mainly attributed to the compositional changes on a localized scale. However, microscopic evidence from imaging and quantification of the chemical changes at this scale due to oxidation is not well documented and the role of oxygen on the crystallization of $Ge_2Sb_2Te_5$ is not well understood yet.

Kooi and coworkers[16] have reported that oxidation of $Ge_2Sb_2Te_5$ significantly affects its crystallization temperature. They showed through in-situ transmission electron microscopy (TEM) studies that a 10 nm film kept in ambient condition for 2 weeks required only 35°C for complete transformation from amorphous to crystalline phase whereas films kept in vacuum required 130°C. They also observed amorphous grain boundaries in the crystalline phase, likely due to formation of amorphous germanium oxide. Of the elements present in the $Ge_2Sb_2Te_5$, Ge is more prone to oxidation owing to its affinity towards oxygen, leading to preferential oxidation of Ge. The remaining film is then rich in Sb and Te. A systematic and quantitative study of chemical changes at varying temperatures preferably at atomic scale is required to unearth issues related to chemical segregation and phase separations in $Ge_2Sb_2Te_5$ with oxygen incorporated.

This letter presents initial results of a systematic microscopic investigation that has been carried out to map and quantify the chemical changes and the compositional segregation in uncapped $Ge_2Sb_2Te_5$ films exposed to atmosphere, as a function of time and temperature.

$Ge_2Sb_2Te_5$ thin films were deposited over amorphous SiNx in Protochips holders by 20 W (DC) power sputtering using a $Ge_2Sb_2Te_5$ target, in 10 mTorr with 10 sccm Ar. The GST films were left uncapped and kept in atmosphere. The target thickness of the deposited film was 30 nm. Deposited and target thicknesses were found to be in close agreement through EELS measurements which yielded a 38 nm SiN membrane thickness, compared to ~ 40 nm specified by the holder manufacturer. Microstructural and microchemical characterization were carried out about four months after deposition, in STEM-HAADF mode employing FEI G2 80-200 chemiSTEM in probe corrected mode with monochromated beam. Chemical distribution of the specimen was determined with four quadrant EDS detectors. Protochips holders (Aduro 300) were used for in-situ heating of the $Ge_2Sb_2Te_5$ film inside the TEM. The Cliff-Lorimer factors (k-factors) were determined from binary GeTe and $Sb_2Te_3$ bulk targets using STEM-XEDS profiles and found to be 3.41 and 1.07 respectively.

Figure 1 shows a representative STEM-HAADF image of a $Ge_2Sb_2Te_5$ thin film of target thickness 30 nm deposited on a Protochips TEM fusion holder. The as-deposited film is comprised of isolated islands with a bimodal size distribution of average sizes ~15 nm and 25 nm. The space between the islands is nearly uniform and ~5 nm in width.

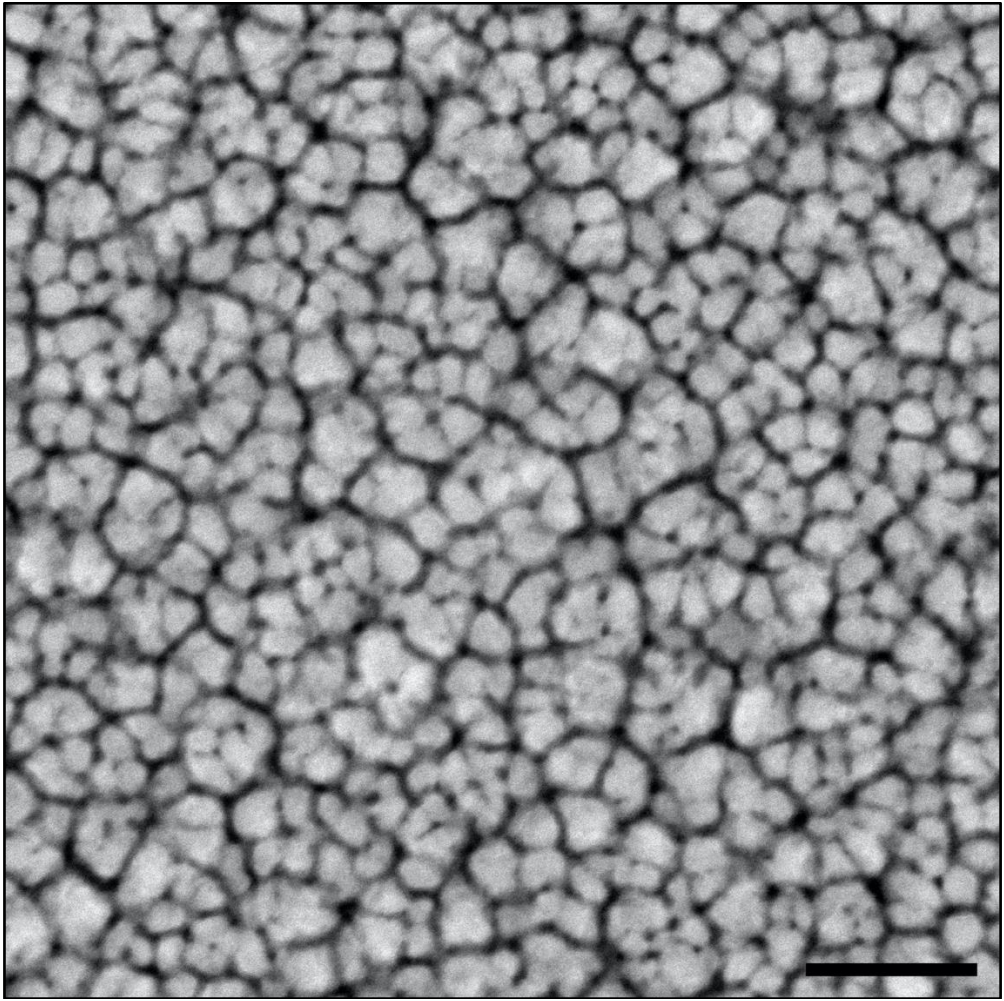

**FIG. 1**. Low-magnification scanning transmission electron microscope high-angle annular dark field (STEM-HAADF) image of $Ge_2Sb_2Te_5$ thin film of 30 nm thickness deposited by sputtering. The morphology shows island like nature of the sputtered film in as-deposited condition. The scale bar is 50 nm.

The generated power spectra (FFT) of the high-resolution HAADF images acquired during in-situ heating are presented in Figure 2. Figure 2(a) is from the as-deposited film and confirms the amorphous nature of the specimen. The power spectrum in Figure 2(b) shows the beginning of the amorphous to crystalline transformation, observed at 130 ºC. The pattern could be indexed as metastable *fcc* phase of $Ge_2Sb_2Te_5$ (ICDD # 00-054-0484). With further increase in temperature, the *fcc* to *hexagonal* transformation was observed and the representative power spectrum at 200 ºC is shown in Figure 2(c). This pattern could be indexed to that of *hexagonal* phase of $Ge_2Sb_2Te_5$ (ICDD#01-082-8882).

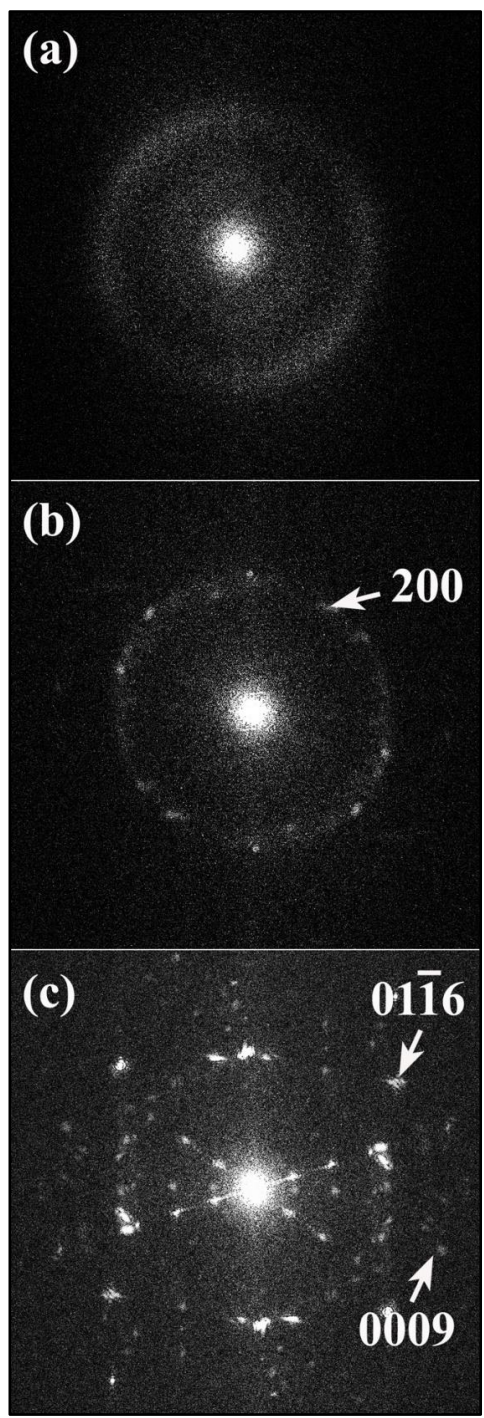

**FIG. 2**. Power spectrum from STEM-HAADF images of (a) as-deposited GST film at room temperature, heated at (b) 130 °C and (c) 200 °C. The phase transformation from amorphous to metastable face-centered cubic (fcc) phase is observed at 130 °C and the second transformation from fcc to a stable hexagonal closed pack (hcp) phase at 200 °C during in-situ heating in the microscope under vacuum.

XEDS analysis of the as-deposited Ge$_2$Sb$_2$Te$_5$ film shows that a significant amount of oxygen is also present (Figure 3). The peaks corresponding to Si and N can be attributed to the SiN$_x$ membrane of the Protochips holder.

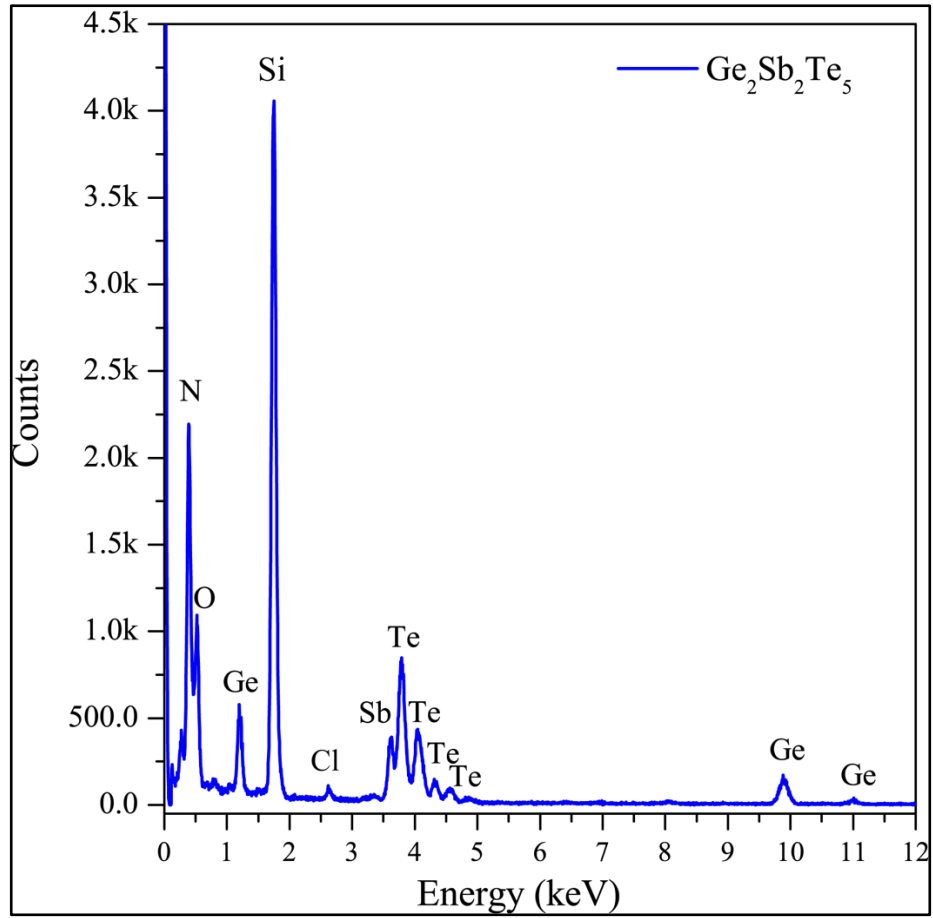

**FIG. 3**. X-ray energy dispersive spectrum of as-deposited 30 nm GeSbTe film showing significant oxygen content.

A series of STEM-XEDS elemental maps acquired in drift-corrected mode from the $Ge_2Sb_2Te_5$ film heated to 200 °C for about 1 hr are shown in Figure 4, showing the distribution of Ge, Sb and Te (at. %). The concentration distribution scales associated with each of the elemental distribution maps show that the islands are almost depleted of Ge, which migrated to the gaps between the islands, whereas Sb and Te are predominantly found in the islands. A composite map of all the elements is depicted in Figure 4(d).

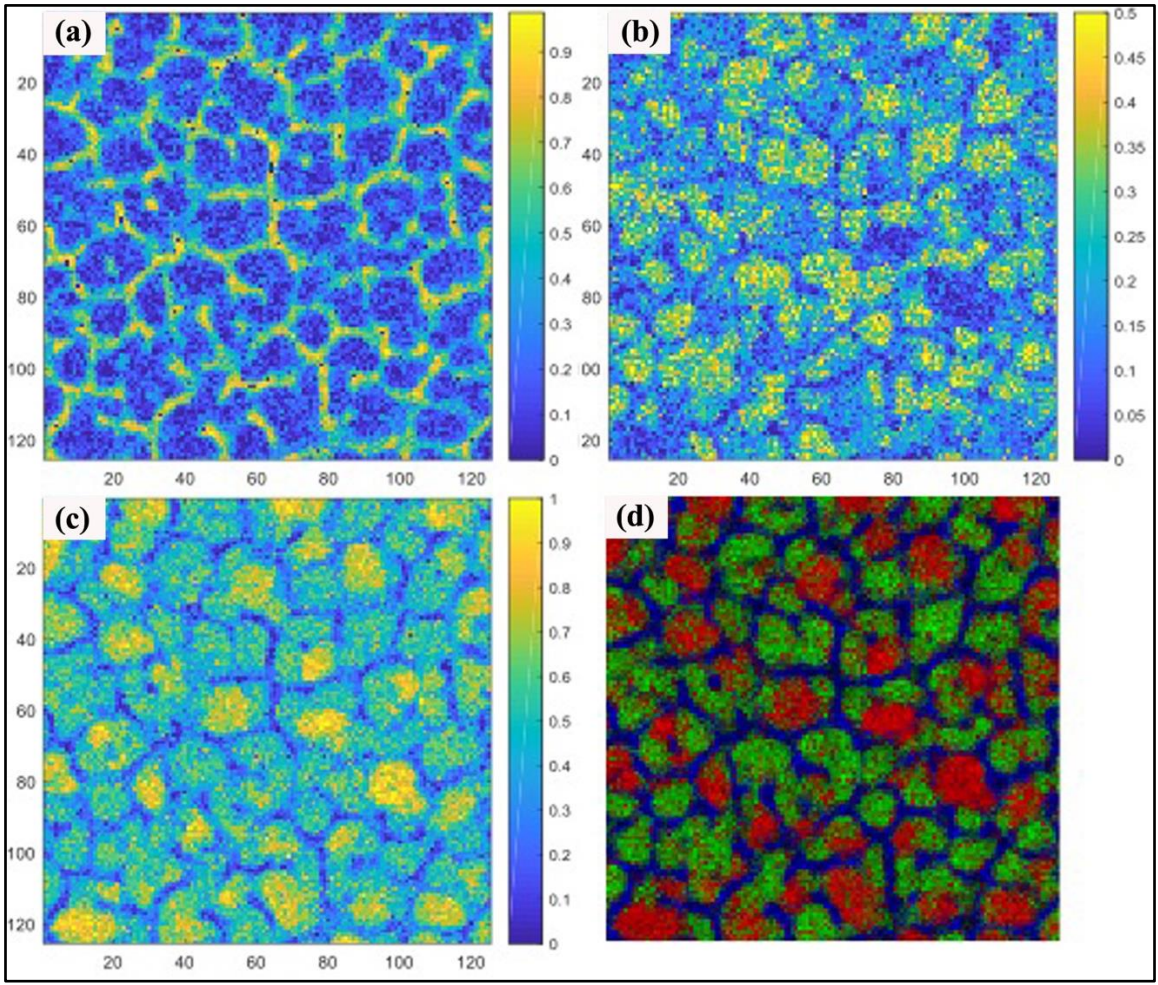

**FIG. 4**. STEM-HAADF chemical mapping of $Ge_2Sb_2Te_5$ thin film at room temperature after heating at 200 °C showing elemental segregation of (a) Ge, (b) Sb, and (c) Te during phase transformation. The scale bar is different for each of the maps. The composite image is shown in (d) where Ge, Sb and Te are indicated in blue, green and red respectively. The presence of oxygen in the gaps of island boundaries and Ge diffusing in the gaps towards oxygen rich areas shows higher affinity of Ge towards O. The field of view for all the images is 125 nm.

To quantify the chemical composition locally, STEM-XEDS elemental line profiles were acquired. The representative STEM-HAADF image and STEM-XEDS elemental line profiles across several islands (shown as a red box) after heat treatment at 200 °C are shown in Figure 5. The morphology is similar to that of the as-deposited sample with appearance of band like contrast in some of the islands. The line profiles further confirm the depletion of Ge in the islands with its redistribution in the inter-island regions and the enrichment of Sb and Te in the islands. The compositional distribution of Sb and Te along the band like morphology regions correspond to $Sb_2Te_3$ composition. In order to investigate the influence of residual oxygen during heat treatment, composite of elemental distribution map including oxygen distribution was also been generated.

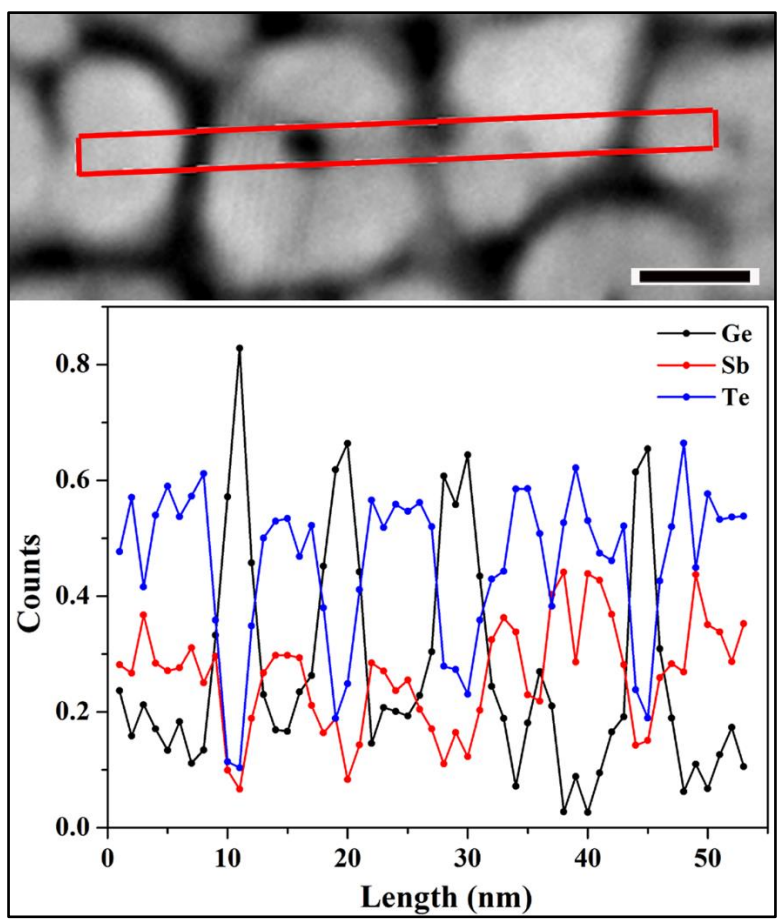

**FIG. 5**. STEM-HAADF image and XEDS line profiles (red line from left to right) of Ge, Sb and Te after heat treatment at 200 ºC. The scale bar for the image is 10 nm.

Figure 6 shows the XEDS profile with composite map of Ge, Sb, Te, and O as inset. The XEDS profile shows that a significant amount of oxygen remained in the film after the heat treatment. The XEDS peaks corresponding to Ge and O are well separated, and their distribution in the composite map is directly interpretable. Interestingly, the distribution of oxygen has been observed in between the islands only. It is important to note that the gaps between the islands are rich in Ge and O whereas the islands are mostly composed of Sb and Te. The composite image also shows that Sb and Te are not homogeneously distributed between islands, with two distinct compositions present (shown as yellow and red islands). The Si in the profile is expected to have originated from the low-temperature deposited, silicon-rich SiNx membrane of the Protochips holders.

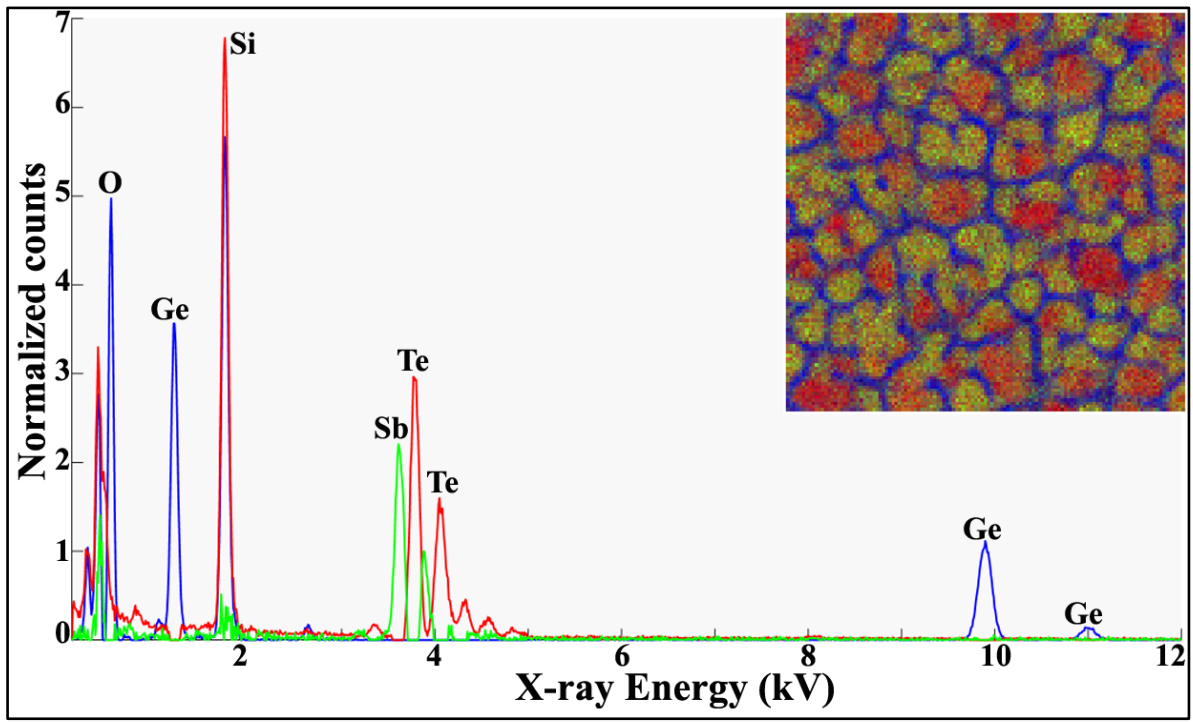

**FIG. 6**. XEDS profile and composite image (inset) of $Ge_2Sb_2Te_5$ including oxygen after heat treatment at 200 °C. The inter islands regions are enriched in Ge and O. The field of view for the composite image is 125 nm.

The accurate composition of the as-deposited and 200 ºC annealed films are calculated based on the k-factor determination from the STEM-XEDS results. The absolute concentration of Ge, Sb and Te in the as-deposited film is determined as 0.272, 0.226 and 0.501 in at. % respectively. This establishes the stoichiometry of the as-deposited film as ~ $Ge_{2.7}Sb_{2.2}Te_5$. Similar studies on the heat-treated film determines the absolute concentration of Ge, Sb, and Te as 0.297, 0.205, and 0.498 in at. % establishing the stoichiometry of the annealed film as ~ $Ge_3Sb_2Te_5$. The chemical composition analysis points out the enrichment of Ge in the as-deposited film itself compared to that of the standard $Ge_2Sb_2Te_5$ target. The Ge concentration increases slightly and the Sb concentration decreases slightly after heat-treatment at 200 ºC whereas the Te concentration remains the same. The presence of oxygen in the inter-island regions, due to exposure of the uncapped film to atmosphere, further promotes the Ge segregation from the islands with the increase in temperature. The changes in elemental compositions shown in the figure 4 indicate that Ge is completely migrated to the inter-island regions and combined with oxygen. It is possible that owing to the non-stoichiometric nature of $SiN_x$, a thin $SiO_2$ layer also formed on the surface of the silicon rich $SiN_x$ membrane, and with an increase in temperature oxygen became available to react with Ge in the film. In the presence of excess oxygen, chemical segregation of $Ge_2Sb_2Te_5$ into GeO and $Sb_2Te_3$ is more viable than crystallization into the *fcc* and subsequent *hexagonal* phase. The bond-dissociation energy for Ge-O (657.5 kJ/mol) is much higher than that of the Ge-Te (396.7 kJ/mol), Te-O (377 kJ/mol) and Sb-O (434 kJ/mol).[17] The formation of Ge-O compound is thermodynamically more favorable than Ge remaining in the GeSbTe cluster or the formation of Sb-O or Te-O compounds. However, this situation may change when $Ge_2Sb_2Te_5$ crystalizes before exposure to oxygen. This chemical segregation occurred in a very localized level and could only be investigated with the aid of TEM. The elemental profiles (cf. figure 5) indicate the formation of $Sb_2Te_3$ in some of the islands where band contrast has been observed.

In summary, as-deposited $Ge_2Sb_2Te_5$ at room temperature, by sputtering of a $Ge_2Sb_2Te_5$ target, uncapped, and subsequently exposed to atmosphere, consists of separate islands, with an overall composition of ~ $Ge_{2.7}Sb_{2.2}Te_{5.2}$. After heating to 200 °C the overall composition becomes ~ $Ge_3Sb_2Te_5$. Ge preferentially interacts with oxygen and migrates towards the inter-islands regions. This chemical segregation has been confirmed through STEM-XEDS and leads to the phase separation of $Ge_2Sb_2Te_5$ films into amorphous Ge-O boundaries and crystalline Sb-Te rich domains. The formation of $Sb_2Te_3$ has also been observed locally. These results are important to achieve improved $Ge_2Sb_2Te_5$ phase-change memory devices especially to help understand the effects of interfaces with silicon dioxide and silicon nitride and of any oxygen content on the electrical and thermal properties of the active phase-change regions.

**Acknowledgments**

This research is supported by NSF under award DMR-1710468. The TEM studies were carried out at CINT, an Office of Science User Facility operated for the U.S. DOE, and in the Materials Characterization Department, Sandia National Laboratories. The GST films were deposited at the Institute of Materials Science and Nanotechnology (UNAM) at Bilkent University, Turkey. The authors thank Dr. Khalid Hattar, Dr. Matthew T Janish and Dr. Ali Gokirmak for helpful discussions. The Sandia National Laboratories are managed and operated by National Technology and Engineering Solutions of Sandia, LLC., a wholly owned subsidiary of Honeywell International, Inc., for the U.S. DOE's NNSA under contract DE-NA-0003525. The views expressed here do not necessarily represent the views of the U.S. DOE or the U.S. Government.

**References**

1. S. Raoux, G. W. Burr, M. J. Breitwisch, C. T. Rettner, Y.-C. Chen, R. M. Shelby, M. Salinga, D. Krebs, S.-H. Chen, H.-L. Lung, C. H. Lam, "Phase-change random access memory: a scalable technology," IBM J. Of Res. and Devel., 52, 4.5 465 (2008).

2. H.-S. Philip Wong, S. Raoux, S.-B. Kim, J. Liang, J. Reifenberg, B. Rajendran, M. Asheghi, and K. E. Goodson, "Phase Change Memory," in *Proceedings of the IEEE*, 98, 12, 2201, (2010).

3. G. W. Burr, M. J. Brightsky, A. Sebastian, H.-Y. Cheng, J.-Y Wu, S. Kim, N. E. Sosa, N. Papandreou, H.-L. Lung, H. Pozidis, E. Eleftheriou, and C. H. Lam, "Recent Progress in Phase-Change Memory Technology," in *IEEE Journal on Emerging and Selected Topics in Circuits and Systems*, vol. 6, no. 2, 146 (2016).

4. J. Hegedus, S. R. Elliott, "Microscopic origin of the fast crystallization ability of Ge-Sb-Te phase-change memory materials", *Nature Materials 7,* 5, 399 (2008).

5. F. Dirisaglik, G. Bakan, Z. Jurado, S. Muneer, M. Akbulut, J. Rarey, L. Sullivan, M. Wennberg, A. King, L. Zhang, "High speed, high temperature electrical characterization of phase change materials: metastable phases, crystallization dynamics, and resistance drift", *Nanoscale 7,* 4, 16625 (2015).

This article has been submitted to Applied Physics Letters. If published, it will be found at https://publishing.aip.org/resources/librarians/products/journals/.6. S. Tripathi, M. Janish, F. Dirisaglik, A. Cywar, Y. Zhu, K. Jungjohann, H. Silva, C. B. Carter, "Phase-Change Materials; the Challenges for TEM," *Microscopy and Microanalysis*, *24*, S1, 1904 (2018).

7. D. Lencer, M. Salinga, M. Wuttig, "Design Rules for Phase-Change Materials in Data Storage Applications," *Adv. Mater*, *23,* 18, 2030 (2011).

8. M. Wuttig, N. Yamada, "Phase-change materials for rewriteable data storage'" *Nature Material*s, *6*, 11, 824 (2007).

9. N. Ciocchini, M. Laudato, M. Boniardi, E. Varesi, P. Fantini, A. L. Lacaita, D. Ielmini, "Bipolar switching in chalcogenide phase change memory," *Scientific Reports* 6, 29162 (2016).

10. P.-Y. Du, J. Y. Wu, T.-H. Hsu, M.-H. Lee, T.-Y. Wang, H.-Y. Cheng, E.-K. Lai, S.-C. Lai, H.-L. Lung, S.-B. Kim, M. J. BrightSky, Y. Zhu, S. Mittal, R. Cheek, S. Raoux, E. A. Joseph, A. Schrott, J. Li, and C. Lam "The impact of melting during reset operation on the reliability of phase change memory", 2012 IEEE *Int. Reliability Physics Symposium*, 6C.2 (2012).

11. Yoon, S.-M.; Choi, K.-J.; Lee, N.-Y.; Lee, S.-Y.; Park, Y.-S.; Yu, B.-G., "Nanoscale observations of the operational failure for phase-change-type nonvolatile memory devices using Ge2Sb2Te5 chalcogenide thin films," *Applied Surface Science 254*, 316 (2007).

12. M. Jang, S. Park, D. Lim, M.-H. Cho, K. Do, D.-H. Ko, H. Sohn, "Phase change behavior in oxygen-incorporated Ge 2 Sb 2 Te 5 films," *Applied Physics Letters 95*, 012102 (2009).

13. E. Gourvest, B. Pelissier, C. Vallée, A. Roule, S. Lhostis, S. Maitrejean, "Impact of oxidation on Ge2Sb2Te5 and GeTe phase-change properties," *Journal of The Electrochemical Society 159*, H373 (2012).

14. L. Yashina, R. Püttner, V. Neudachina, T. Zyubina, V. Shtanov, M. Poygin, "X-ray photoelectron studies of clean and oxidized α-GeTe (111) surfaces," *Journal of Applied Physics, 103*, 094909 (2008).

15. C. Rivera-Rodrıguez, E. Prokhorov, G. Trapaga, E. Morales-Sanchez, M. Hernandez-Landaverde, Y. Kovalenko, J. Gonzalez-Hernandez, "Mechanism of crystallization of oxygen-doped amorphous Ge1Sb2Te4 thin films," *Journal of Applied Physics 96*, 2, 1040 (2004).

16. B. Kooi, W. Groot, J. T. M. De Hosson, "In situ transmission electron microscopy study of the crystallization of Ge2Sb2Te5," *Journal of Applied Physics 95,* 3, 924 (2004).

17. Luo, Y.-R., *Handbook of bond dissociation energies in organic compounds*. CRC press: 2002.
12/12